\documentclass[prd,preprint,superscriptaddress,showpacs,byrevtex]{revtex4}

\usepackage{epsfig}

\newcommand{\be}{\begin{equation}}
\newcommand{\ee}{\end{equation}}
\newcommand{\bea}{\begin{eqnarray}}
\newcommand{\eea}{\end{eqnarray}}

\begin{document}

\title{Soft-Gluon Production Due to a Gluon Loop in a
Constant Chromo-Electric Background Field}

\author{Gouranga C. Nayak } \email{nayak@insti.physics.sunysb.edu}
\affiliation{
C. N. Yang Institute for Theoretical Physics, Stony Brook University, 
SUNY, Stony Brook, NY 11794-3840, USA }

\author{Peter van Nieuwenhuizen} \email{vannieu@insti.physics.sunysb.edu}
\affiliation{
C. N. Yang Institute for Theoretical Physics, Stony Brook University, 
SUNY, Stony Brook, NY 11794-3840, USA }

\date{\today}
\begin{abstract} 

We obtain an exact result for the soft gluon production and its $p_T$ distribution 
due to a gluon loop in a constant chromo-electric background field $E^a$ 
with arbitrary color. 
Unlike Schwinger's result for $e^+e^-$ pair production in QED
which depends only on one gauge invariant quantity,
the electric field $E$, we find that the $p_T$ distribution of the 
gluons depend on two gauge invariant quantities, 
$E^aE^a$ and $[d_{abc}E^aE^bE^c]^2$.

\end{abstract} 

\maketitle
\newpage

Although hard gluon production at high energy colliders
is computed by using perturbative quantum chromodynamics (pQCD),
it is not possible to calculate soft gluon 
production by using perturbation theory
because the coupling constant becomes large at low energy. In this paper
we study non-perturbative soft gluon production due to vacuum polarization in a
constant chromo-electric field by applying the background field method to QCD
with gauge group SU(3). While $e^+e^-$ production 
from a constant electromagnetic
field in spinor QED has been calculated long ago by Euler and Heisenberg
\cite{euler}, and Schwinger \cite{schw}, and by Weisskopf \cite{wisk} for
scalar QED,
there is of course no direct photon production
in QED as photons do not interact with a classical electromagnetic
field. In the case of QCD, gluons do interact with a background chromofield
and gluons are produced. As we want to study the $p_T$ distribution
of gluons produced at collider experiments, we shall not employ Schwinger's 
proper time method but, rather, we will directly evaluate 
the path integral and obtain the partition function
which gives both the $p_T$ distribution and the total gluon production.

For the interpretation of experiments which probe the quark-gluon plasma
at high-energy 
large-hadron colliders such as RHIC (Au-Au collisions at $\sqrt s$ = 200 GeV)
\cite{rhic} and LHC (Pb-Pb collisions at $\sqrt s$ = 5.5 TeV) \cite{lhc} it 
might be necessary
to know the $p_T$ spectrum of soft gluons produced by a chromo field.
The physical picture behind this calculation is that two heavy nuclei collide
and then move apart, creating a classical chromo field in between
\cite{all,all1}. 
We consider here a constant chromo-electric field along the beam direction.
Subsequently this constant chromo-electric field 
breaks up into quark anti-quark pairs and gluons. 
The quark anti-quark production
has been calculated before \cite{yildiz1} and the corresponding $p_T$ distribution 
has been approximated by WKB methods \cite{wkb}. Also the total gluon production rate
by a covariantly constant field ($dN/d^4x$) has been calculated for SU(2) 
in \cite{sav,schd} and for SU(3) in \cite{yildiz}. 
The $p_T$ distribution ($dN/d^4xd^2p_T$) of gluon production
due to a constant chromo-electric field which we present in this article is new,
as far as we know, and we consider the realistic case of SU(3). 

We obtain the
following formula for the number of non-perturbative soft gluons produced 
per unit time and per unit volume and per unit transverse momentum from a given 
constant chromo-electric field $E^a$
\bea
\frac{dN_{gg}}{dt d^3x d^2p_T}~
=~\frac{1}{4\pi^3} ~~ \sum_{j=1}^3 ~
~|g\lambda_j|~{\rm ln}[1~+~e^{-\frac{ \pi p_T^2}{|g\lambda_j|}}].
\label{1}
\eea
This result is gauge invariant because $|\lambda_1|$, $|\lambda_2|$, and $|\lambda_3|$ 
are the positive square root of the following gauge invariant real positive quantities
\bea
&&~\lambda_1^2~=~\frac{C_1}{2}~[1-{\rm cos}~ \theta],  \nonumber \\
&&~\lambda_2^2~=~\frac{C_1}{2}~[1+{\rm cos}(\frac{\pi}{3}-\theta)],  \nonumber \\
&&~\lambda_3^2~=~\frac{C_1}{2}~[1+{\rm cos}(\frac{\pi}{3}+\theta)],
\label{lm}
\eea
where $\theta$ is real and is given by
\bea
cos^3\theta~=-1+6C_2/C_1^3.
\label{theta}
\eea
They depend only on the Casimir invariants for SU(3)
\bea
C_1~=~E^aE^a, ~~~~~~~~~~~~~~~~~~~~~~~
C_2~=~[d_{abc}E^aE^bE^c]^2,
\label{casm}
\eea
where $a,~b,~c$ ~=~1,...,8 are the color 
indices of the adjoint representation of the gauge group 
SU(3). Note that $\theta$ is real because $C_1^3-3C_2~\ge~0$.

This can be contrasted with the corresponding formula for massless 
fermion pair production 
\bea
\frac{dN_{e^+e^-}}{dt d^3x d^2p_T}~=~\frac{-|eE|}{4\pi^3} ~ 
{\rm ln}[1~-~e^{-\frac{\pi p_T^2}{|eE|}}].
\label{7}
\eea
The result for the $p_T$ distribution in (\ref{7})
was obtained by WKB methods in \cite{wkb} and integration over $p_T$
reproduces Schwinger's result for total production rate $dN/d^4x$ \cite{schw}. 
(The 
first term in the expansion of the logarithm of 
the total production rate $dN/d^4x$ was already obtained
by Heisenberg and Euler
\cite{euler}. For an excellent review on Euler-Heisenberg actions, see \cite{dune}).
In our result the symmetric tensor $d_{abc}$ appears. Hence the extension
of Schwinger's formula for $e^+e^-$ pair production  
to gluon production is not straight forward. We now present a derivation.

The Lagrangian density in the background field method of QCD with covariant background
gauge fixing term in Feynman-'t Hooft gauge is given by \cite{thooft}
\be
{\cal{L}}_{gluon}~=~
\frac{1}{2}~Q^{\mu a}~M^{ab}_{\mu \nu}[A]~Q^{\nu b}
\label{lag}
\ee
where
\bea
M_{\mu \nu}^{ab}[A]~=~\eta_{\mu \nu}[D_\rho(A)D^\rho(A)]^{ab}~-~2gf^{abc}F_{\mu \nu}^c 
\eea
with $\eta_{\mu \nu}~=~(-1,+1,+1,+1)$ the Minkowski metric. 
The corresponding ghost Lagrangian density is given by
\be
{\cal{L}}_{ghost} ~=~ \overline{\chi}^a D_{\mu}^{ab}[A]
D^{\mu ,bc}[A+Q]\chi^c~=~\overline{\chi}^a~K^{ab}[A,Q]~\chi^b 
\label{ghos1}
\ee
where Q, A and $\chi$ are the gluon quantum field, the background field,
and the ghost field, respectively, and
$D_\mu[A]Q_\nu^a~=~[\partial_\mu Q_\nu^a~+~gf^{abc}A_\mu^bQ_\nu^c]$. 

The one-loop effective action for a gluon loop in a background field $A_{\mu}^a$ is given by
\bea
S^{(1)}_{gluon}~=~-i {\rm ln}(Det~M)^{-1/2}~=~\frac{i}{2}~Tr[
{\rm ln}M[A]~H- {\rm ln}M[0]~H]
\label{efgl}
\eea
where $H^{\mu \nu}~=~\eta^{\mu \nu}$. 
We added the matrices $H^{\mu \nu}$ since this will allow us to factorize
the trace over Lorentz indices. 
The trace Tr contains an integration over $d^4x$ and 
a sum over color and Lorentz indices. We replace
the logarithm by using the exponential representation
\bea
S^{(1)}_{gluon}~=~\frac{i}{2} ~Tr \int_0^\infty~\frac{ds}{s} [
e^{is~(M[0]~H+i\epsilon)}~ -e^{is~(M[A]~H+i\epsilon)}].
\eea
We shall say more about the s-integration contour later.
We assume that 
the electric field is along the z-axis (the beam direction) 
and we choose the gauge $A_0^a=0$
so that
$A_3^a~=~-E^a \hat{x^0}$.
The color indices  (a=1,....8) are arbitrary. 
Since $\Lambda^{ab}~=~if^{abc}E^c$
is hermitian and anti-symmetric, its eigenvalues
are real and come in pairs $(\lambda,~-\lambda)$. 
We will show later that the matrix $\Lambda^{ab}$
has six non vanishing eigenvalues. So after diagonalization it reads
\bea
\Lambda_{d}^{ab}~=~( \lambda_1, -\lambda_1, \lambda_2,-\lambda_2, \lambda_3, -\lambda_3,0,0).
\label{eigen}
\eea
We replace the derivative $\partial_\mu$ by $ip_\mu$ satisfying 
$[p_\mu, x^\nu]=-i\delta_\mu^\nu$ and obtain
\bea
M[A]_{\mu \rho}^{ab}~\eta^{\rho \nu}~=~ \delta_\mu^\nu~[ ({\hat{p_0}}^2 -{\hat{p_T}}^2 
-{\hat{p_3}}^2)\delta^{ab} 
~ -~2g\Lambda^{ab} {\hat{x^0}} {\hat{p_3}}~-~ 
g^2{(\Lambda^2)}^{ab} {\hat{x^0}}^2 ] ~+~2ig\Lambda^{ab}~S_\mu^{~\nu}
\label{map}
\eea
where $p_T~=~\sqrt{p_1^2+p_2^2}$ is the transverse momentum of the gluons (transverse
to the electric field direction) and
\bea
S_\mu^{~\nu}~=~
\left[ \begin{array}{cccc}
0 & 0 & 0 & 1  \\
0 & 0 & 0 & 0 \\
0 & 0 & 0 & 0\\
1 & 0 & 0 & 0
\end{array} \right].
\eea
To reduce this problem to one harmonic oscillator, we make 
the usual similarity transformation \cite{itz}. We also make
a similarity transformation in group space which 
diagonalizes the matrices $\Lambda$
\bea
M[A]_{\mu \rho}^{ab}~\eta^{\rho \nu}~=~ {\big \{  e^{ip^3p_0/g\Lambda_d}~
\big[ ({\hat{p_0}}^2 -{\hat{p_T}}^2 ~-~g^2{(\Lambda_d^2)} {\hat{x^0}}^2)
~\delta_\mu^\nu  ~
+~2ig\Lambda_d~S_\mu^{~\nu} \big]~ e^{-ip^3p_0/g\Lambda_d} \big \}}^{ab}.
\label{map1}
\eea
Recalling that $\Lambda^{ab}$
has two vanishing eigenvalues, we should read this expression as follows: for a=b=1,..6
each $\Lambda_d$ denotes an eigenvalue $\pm \lambda_1,~ \pm \lambda_2~ \pm \lambda_3$,
while for a=b=7,8 one finds an 
expression which is independent of $A_\mu^a$ and which we omit because it cancels 
against the contributions from $M[0]$.

The trace over Lorentz indices factorizes because $\delta_\mu^{~\nu}$ 
commutes with $S_\mu^{~\nu}$ and we obtain
\bea
&&S^{(1)}_{gluon}~=~\frac{-i}{2}~\int_0^\infty~\frac{ds}{s}~\sum_{j=1}^6~ tr \Big[ 
e^{ip^3p_0/g\lambda_j}~
e^{is({\hat{p_0}}^2 -{\hat{p_T}}^2 ~-~g^2\lambda_j^2 {\hat{x_0}}^2 +i\epsilon)}~
e^{-ip^3p_0/g\lambda_j} \nonumber \\
&&~[2+2cosh2sg\lambda_j] ~-~4e^{is({\hat{p_0}}^2 -{\hat{p_T}}^2 +i\epsilon)} \Big]~
\label{trhg}
\eea
The trace tr denotes integral over a complete set of x eigenstates.
From here we proceed as in QED \cite{itz}.
We add complete sets of p eigenstates, and
obtain 
\bea
S^{(1)}_{gluon}~=~\frac{-i}{2}~\int_0^\infty~\frac{ds}{s}~\sum_{j=1}^6~ \frac{1}{(2\pi)^3}~\int d^4x~\int
d^2p_T~
e^{-isp_T^2 ~-~s\epsilon}[~|g\lambda_j| \frac{1+cosh2sg\lambda_j}{sinhs|g\lambda_j|}-\frac{2}{s}~].
\label{trg}
\eea

The one-loop effective action for the ghost in the background field $A_\mu^a$
is given by
\bea
S^{(1)}_{ghost}~=~-i {\rm ln}(Det~K)~=~-i~Tr \int_0^\infty~\frac{ds}{s} [
e^{is~[K[0]+i\epsilon]}~ -e^{is~[K[A]+i\epsilon]}]
\eea
where $K^{ab}[A]$ is given by (\ref{ghos1}). Since there is now no trace 
over $\mu,~ \nu$ but also no square root of the ghost determinant, 
we find an overall factor $-1/2$. In addition there is no term 
$f^{abc}F_{\mu \nu}^c$ in $K[A]$ so that the term $cosh2sg\lambda_j$ is absent.
We obtain then
\bea
S^{(1)}_{ghost}~=~\frac{i}{2}~\int_0^\infty~\frac{ds}{s}~\sum_{j=1}^6~ \frac{1}{(2\pi)^3}~
\int d^4x~\int d^2p_T~
e^{-isp_T^2 ~-~s\epsilon}[ |g\lambda_j| \frac{1}{sinhs|g\lambda_j|}-\frac{1}{s}].
\label{trgh}
\eea
Adding the effective action due 
to a gluon and a ghost loop we find for the one-loop effective action
\bea
S^{(1)}~=~\frac{-i}{2}~\int_0^\infty~\frac{ds}{s}~\sum_{j=1}^6~ \frac{1}{(2\pi)^3}~\int d^4x~
\int d^2p_T~
e^{-isp_T^2 ~-~s\epsilon}[|g\lambda_j| \frac{cosh2sg\lambda_j}{sinhs|g\lambda_j|}-\frac{1}{s}].
\label{tea}
\eea
The s-integral at fixed $p_T$ is convergent at s $\rightarrow$ 0, but integration 
over $p_T$ yields an extra factor 1/s, and charge 
renormalization cures this ultraviolet problem by substracting also
the term linear in $s$ in the expansion of
$\frac{cosh2s|g\lambda_j|}{sinhs|g\lambda_j|}$.
In fact the term linear in s yields the beta function
\cite{sav, schd}. 

A more serious problem occurs for s $\rightarrow \infty$.
Obviously, the integral over s diverges, indicating the presence of infrared
divergences. Although such divergences in Feynman graphs with on-shell
external gluons 
are well known, the full non-perturbative treatment of these divergences
seems complicated. However, we are only interested in the imaginary part of the
effective action, since it gives us the gluon production rate, and hence
we proceed similarly to Yildiz and Cox (see eq. 22 in \cite{yildiz}).
We use the well-known expansion
\be
1/sinhx~=~\frac{1}{x}~+~2x~\sum_{n=1}^\infty~\frac{(-1)^n}{x^2+n^2\pi^2}
\label{sinh}
\ee
and then we formally replace $s$ by $-is$ (as first advocated by Schwinger in QED
where no infrared problems are present).
The integral is now real, except for half-circles around the poles
at $s|g\lambda_j|~=~-in\pi$ for n=1,2,3... The term $1/x$ in (\ref{sinh}) cancels
against the term $1/s$ in (\ref{tea}).
This yields probability for gluon production per unit time and per unit volume
\bea
W_{gluon}~=~2ImS^{(1)}~=~\frac{1}{(2\pi)^3} ~\int d^2p_T~ \sum_{n=1}^\infty~(-1)^{n+1}~\sum_{j=1}^6~
\frac{|g\lambda_j|}{n}~
~e^{[-\frac{n \pi p_T^2}{|g\lambda_j|}]}. 
\label{gli6}
\eea

All that is left is to determine the eigenvalues 
$\lambda_j$ (j = 1,...,8) of the matrix $if^{abc}E^c$. 
First we evaluate the determinant of the 8$\times$8 matrix
$[f^{abc}E^c ~-~\lambda~\delta^{ab}]$ and find
\bea
Det[f^{abc}E^c ~-~\lambda~\delta^{ab}]= &&\lambda^2~[\lambda^6~+~A\lambda^4~+~B\lambda^2~+~C]  \nonumber \\
&&~
=\lambda^2(\lambda^2+\lambda_1^2)~ (\lambda^2+\lambda_2^2)~ (\lambda^2+\lambda_3^2). 
\label{eg1}
\eea
Because $E^b$ is an eigenvector of the matrix $if^{abc}E^c$, and because eigenvalues 
come in pairs $(\lambda_j,~ -\lambda_j)$, there are two eigenvalues zero, which explains
the factor $\lambda^2$. 
The gauge invariant quantities A, B, and C in the above equation 
can only depend on the Casimir invariants in (\ref{casm}) and we find
\bea
A~=~\frac{3}{2}E^aE^a, ~~~ B~=~\frac{A^2}{4}, ~~~~
C~=~~\frac{1}{16}~\big([E^aE^a]^3 ~-~3~[d_{abc}~E^aE^bE^c]^2\big),
\label{vl}
\eea
where $d_{abc}$ is the symmetric invariant tensor in the 
adjoint representations of SU(3). It follows that the three eigenvalues satisfy
\bea
&& \lambda_1^2~+~\lambda_2^2~+~\lambda_3^2~=~A \nonumber \\
&& \lambda_1^2\lambda_2^2 ~+~\lambda_2^2\lambda_3^2~+~\lambda_3^2\lambda_1^2~=~B \nonumber \\
&& \lambda_1^2\lambda_2^2\lambda_3^2~=~C,
\eea
the solution of which is given by eq. (\ref{lm}).

In this letter we have derived the $p_T$ distribution of the 
soft gluons produced by the vacuum polarization due to a gluon loop
in a constant chromo-electric field.
We have used the background field method of QCD with gauge group SU(3). 
For application at RHIC and LHC we have constructed the
$p_T$ distribution of the gluon production.
We find that, unlike the case in QED 
where the $e^+e^-$ pair production rate depends on 
one gauge invariant quantity $|E|$, the $p_T$ distribution of
the gluon production rate depends on two gauge 
invariant quantities, $E^aE^a$ and $[d_{abc}E^aE^bE^c]^2$. We intend to 
add a chromo-magnetic field later.

\acknowledgments

We thank Fred Cooper, Gerald Dunne, Martin Rocek, Jack Smith,
George Sterman and William Weisberger for discussions. This work
was supported in part by the National Science Foundation, grants
PHY-0098527, PHY-0354776 and PHY-0354822.

\end{document}